\documentclass[final,5p,times]{elsarticle}

\usepackage{amssymb}
\usepackage{amsmath}
\usepackage{lipsum}
\usepackage{booktabs}
\usepackage{xcolor}

\journal{Nuclear Instruments and Methods in Physics Research Section A}

\begin{document}

\begin{frontmatter}



\title{Improving Optics Control and Measurement at RHIC}


\author[msu]{William Fung}
\author[msu]{Yue Hao}
\address[msu]{Department of Physics and Astronomy, Michigan State University, East Lansing, MI 48824, USA}

\author[bnl]{Xiaofeng Gu}
\author[bnl]{Guillaume Robert-Demolaize}
\address[bnl]{Brookhaven National Laboratory, Upton, NY 11973, USA}

\begin{abstract}

Maximizing luminosity requires precise control of the optics function at the interaction point (IP), implying that the location ($s^*$) of the beta function’s minimum value ($\beta^*$) must be moved to the collision location ($s_{IP}$) as much as possible. Accurate optics measurements and reliable control of $s^*$ in both planes are therefore essential for optimal collider performance. During Relativistic Heavy Ion Collider (RHIC) operations in 2024, measurements indicate an average horizontal beta beat of approximately $20\%$ at IP8, accompanied by measurement variation in measured $s^*$ in both planes. In this paper, a sensitivity-matrix-based optics correction scheme is demonstrated to effectively steer the optics toward desired targets using power supply currents of IR quadrupoles in the 8 o’clock interaction region (IR8). In addition, a method for measuring linear optics based on the one-turn map within the interaction regions is developed and systematically compared with established optics measurement methods used in RHIC operations. A comprehensive error analysis is performed for all measurement methods considered. Through these methods, a consistent reduction of $10\%$ beta beat is achieved by moving $s^*_x$ as well as a significant improvement in the reproducibility of $s^*$ measurements in both planes. The techniques demonstrated here will be further developed to support linear optics analysis and control of the future Electron-Ion Collider project.
\end{abstract}




\end{frontmatter}




\section{Introduction}
\label{introduction}

The luminosity is the figure of merit for delivering high-quality physics measurements at the STAR (IR6) and sPHENIX (IR8) experiments at the Relativistic Heavy Ion Collider (RHIC). The beam optics near each interaction region (IR) must be accurately controlled so that both colliding beams attain their desired transverse beam sizes at the designed longitudinal position relative to the detector location. The IR optics are commonly characterized by the minimum of the beta function ($\beta^* = \beta^*_{x,y}$) located at longitudinal position $s^* = s^*_{x,y}$ at each IR \cite{lum}. Therefore, precise measurements and control of the optics, especially of $s^*_x \text{ and } s^*_y$, are crucial to maximizing the performance of the collider.

In order to locally adjust or correct the optics within an IR of a collider ring without perturbing the global optics, a model-based tuning scheme utilizing the Jacobian solver in MAD-X (commonly referred to as MAD-X matching) is often employed \cite{madxmanual, madxmatching}. At RHIC, this approach has been successfully applied in several contexts, including matching the dispersion derivatives at the Siberian snakes to preserve spin polarization \cite{snakes}. However, due to the inherently nonlinear nature of the solver, the resulting magnet strengths can exhibit significant variation between successive iterations. Such fluctuations pose challenges for online optimization, where large, iterative changes in magnet settings are particularly susceptible to magnet hysteresis effects. As an alternative, this paper considers the use of an optics response matrix, or sensitivity matrix $\boldsymbol{B}$, which enables a linearized approach to optics correction with improved control over changes in magnet strengths \cite{Minty}.

A wide range of techniques have been developed to measure the beta function and minimize discrepancies between measured and model optics (commonly referred to as beta-beat) using beam position monitors (BPMs), as reviewed in \cite{review}. Beta-from-amplitude methods such as Harmonic Analysis (HA) \cite{Castro} and Model Independent Methods (MIA) \cite{MIA} provide rapid and convenient means for optics measurements. However, these approaches are susceptible to systematic errors originating from BPM calibration and noise. Beta-from-phase-advance methods, such as 3-BPM \cite{Castro} and N-BPM methods \cite{NBPM} demonstrate improved robustness against BPM errors and eliminate calibration errors. While the N-BPM is more computationally intensive, various advancements have been developed to improve its computational efficiency \cite{NBPM2}. 

\begin{figure}[htbp]
    \centering
    \includegraphics*[width=1\columnwidth]{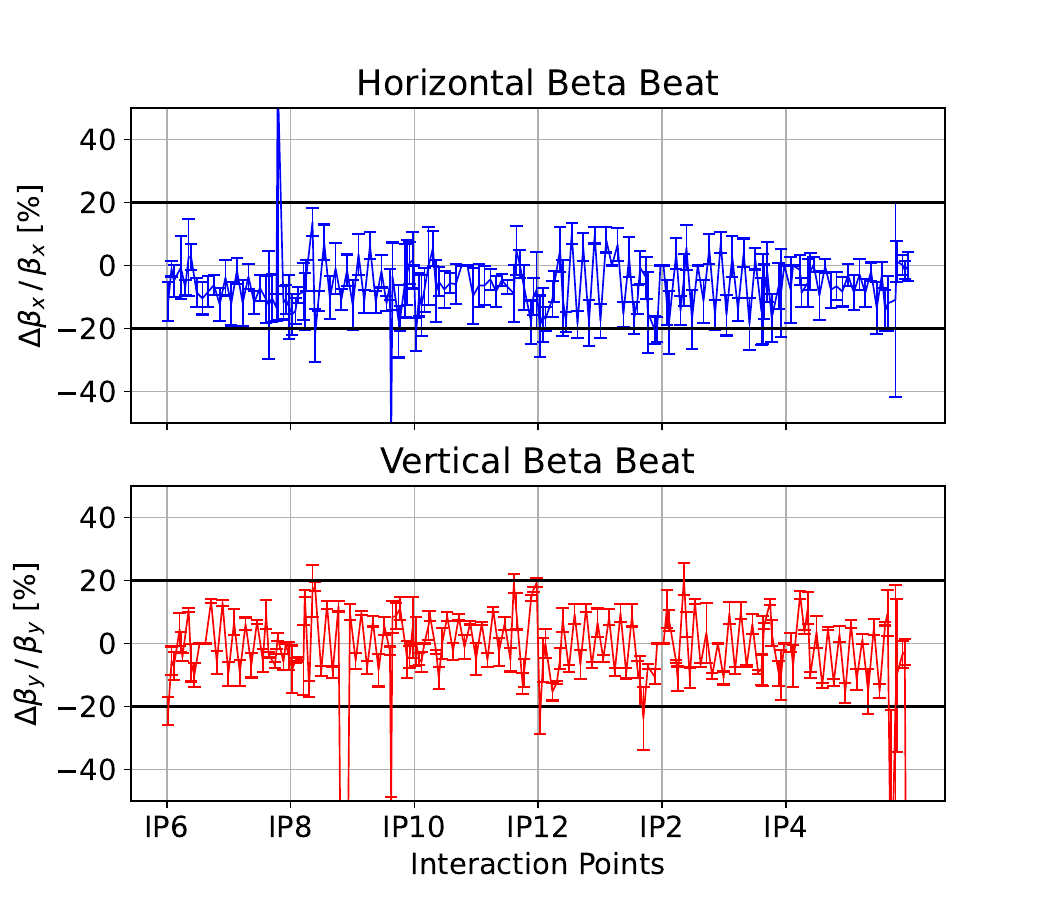}
    \caption{\label{fig:RHIC_beat} Measured horizontal (blue) and vertical (red) beta beat from curve fit method vs BPM number. The solid black lines indicate beta beat $\pm 20 \%$. Error bars were calculated over $1\sigma$ to be around $5\%$ and $3\%$ for horizontal and vertical axis respectively.}
\end{figure}

Several experimental studies in 2024 were carried out using a proton-proton collision lattice at RHIC with a beam energy of $E = 100~\mathrm{GeV}$ and $\beta^* \approx 0.9~\mathrm{m}$ at IP6 and IP8, following the lattice configuration described in \cite{Luo_RHIC_DynamicAperture_2009}. Six initial turn-by-turn (TBT) datasets were then acquired without applying quadrupole modifications to evaluate the robustness of the current optics measurement program in the RHIC control room. The mean and sample standard deviations of the measured beta functions were interpreted as the beta beat and the corresponding measurement uncertainties, respectively, as shown in Figure~\ref{fig:RHIC_beat}. In this work, the beta beat is defined as:

\begin{equation}\label{beta beat}
    \Delta\beta/\beta = \beta \text{ beat} = 100 \times \frac{\beta_{model} - \beta_{meas}}{\beta_{model}} [\%]
\end{equation}

\noindent where $\beta_{meas}$ denotes the measured beta function (in Figure~\ref{fig:RHIC_beat} obtained using the RHIC optics program, R-OP), and $\beta_{model}$ represents the model value from MAD-X. 

The transverse (horizontal and vertical) beta beat at each BPM was found to be within approximately $\pm20\%$. Each BPM's beta beat was calculated using Equation~(\ref{mean abs val}), where $N$ denotes the number of TBT datasets used. The absolute value is taken to extract the magnitude of disagreement for beta measurements compared to the model. The error bars, calculated at the $1\sigma$ level, indicate increased measurement variation compared with previous RHIC analysis \cite{ICARHIC}. Unless otherwise noted, all error bars presented in this paper correspond to $1\sigma$ uncertainties.

\begin{equation}\label{mean abs val}
    \bar{\beta^*}_{abs}\text{ beat} = \frac{1}{N}\sum_{i=1}^{N} \left| \frac{\beta_{i, model} - \beta_{i, meas}}{\beta_{i, model}} \right|
\end{equation}

The measured beta functions from the TBT data were further used to extract the $\beta^*$ beat and the longitudinal shift of the beta-function minimum, $s^*$, at IP8. The mean absolute value of the $\beta^*$ beat was found using Equation~(\ref{mean abs val}), where $N$ denotes the number of TBT datasets used. The mean of the $s^*$ difference, as well as their standard deviations are shown in Table~\ref{R_IP8_Metrics}. The $s^*$ difference will be defined in this paper as:

\begin{equation}\label{s diff}
    \bar{s}^* \text{ difference} = (s_{meas}^* - s_{model}^*) [m]
\end{equation}

In Table~\ref{R_IP8_Metrics}, the model value of the waist location ($s_{model}^*$) is zero. Throughout the paper, a shift of the waist location is denoted by $\Delta s^*$ movement, and $\Delta s^* = s_{model}^*$. For the initial unshifted optics, $\Delta s^*=0$ and $s^*_{model}=0$. When a waist shift is intentionally applied through adjustments computed from $\boldsymbol{B}$, the model waist value $s^*_{model}$ represents the expected waist location of the waist after the requested displacement. 

This analysis reveals a horizontal $\beta^*$ beat of approximately $20\%$ at IP8, together with significant measurement variation in the measured $s^*$ values in both transverse planes. The observed variability highlights limitations in the reproducibility of optics measurements obtained with R-OP when multiple TBT datasets are analyzed. Given that the sPHENIX detector provides vertex detection within $\pm 10~\mathrm{cm}$ \cite{sPHENIX}, improved accuracy and consistency in $s^*$ measurements and corrections are expected to directly enhance luminosity optimization at RHIC.

\begin{table*}[!t]
\caption{\label{R_IP8_Metrics} The mean absolute value of $\beta^*$ beat, mean $s^*$ difference, and sample standard deviations were taken over six TBT measurements with $\Delta s^* = 0$ using R-OP.}
\centering
\begin{tabular}{c
            @{\hspace{12pt}} r@{\,\ensuremath{\pm}\,}l
            @{\hspace{24pt}} r@{\,\ensuremath{\pm}\,}l}
\toprule
R-OP &
\multicolumn{2}{c}{$\bar\beta^*_{abs}$ Beat [\%]} &
\multicolumn{2}{c}{$\bar s^*$ Difference [m]} \\
\midrule
Horizontal & 18 & 7 & 0.0 & 0.4 \\
Vertical   &  7 & 5 & 0.1 & 0.2 \\
\bottomrule
\end{tabular}
\end{table*}


Beta-from-amplitude methods such as R-OP rely on lattice models such as MAD-X to infer the beta function. The model dependency may unintentionally bias results toward design optics and reduce sensitivity to genuine deviations in the machine. Motivated by this limitation, a model-independent method for measuring optics using the one-turn map at the IRs is discussed and compared to various established model-dependent methods.  Previous studies have employed the one-turn map to measure global coupling at RHIC \cite{coupling}; however, those analyses relied on lattice models when magnets were present within the measurement region, rendering them partially model dependent.

The remainder of this paper is organized as follows. Section~\ref{sec: Optics tuning} introduces the proposed algorithm for adjusting $s^*$ using a sensitivity-matrix-based approach. The corresponding current changes are converted to magnet strengths and applied in the RHIC control room. Following excitation of the betatron motion using a fast kicker, TBT data is acquired, preprocessed, and analyzed using the optics measurement techniques described in Section~\ref{sec: Optics Measurement Methods} to extract $\beta^*$ and $s^*$. Finally, Section~\ref{sec: Experimental Results} presents the experimental results and provides a detailed comparison of the different optics measurement methods.

\section{\label{sec: Optics tuning}  Optics tuning}

To achieve better control of the optics at IR8, an approach based on the sensitivity matrix $\boldsymbol{B}$ \cite{Minty} was utilized.  $\boldsymbol{B}$ represents an approximately linear relationship between changes in magnet strengths $\Delta \boldsymbol{K}$ and in optics $\Delta \boldsymbol{O}$.  Such a formulation may request combinations of magnet settings that are not physically realizable due to hardware constraints imposed by the power supply (PS) configuration \cite{RHICConfigManual}.  
For instance, in the interaction regions IR6 and IR8 of RHIC,  quadrupoles 4-6 on both sides of the IP are powered by a common current supply. 
Therefore, the optimization variables and constraints were defined in terms of PS currents $\boldsymbol{I}$ rather than the magnet strengths. Accordingly, the sensitivity matrix $\boldsymbol{B}$ is defined here as the linear relation between the change in the power supply currents $\Delta \boldsymbol{I}$ and the change in optics parameters $\Delta \boldsymbol{O}$: 

\begin{equation}\label{sensitivity relation}
    \Delta \boldsymbol{O} = \boldsymbol{B} \Delta \boldsymbol{I}
\end{equation}

The vector $\boldsymbol{I}$ consists of seventeen PS currents that affect seventeen insertion quadrupoles within IR8. The vector $\boldsymbol{O}$ includes thirteen optics parameters, including transverse $s^*, \beta^*$, and horizontal dispersion at IP, as well as optics quantities at the upstream end of IR8 (between insertion and arc regions): transverse $\beta, \alpha, \mu$, horizontal dispersion, and slope of the horizontal dispersion. The optics parameters in $ \boldsymbol{O}$ were chosen to match the design optics at IP and the arc optics at the ends of IR8.



The sensitivity matrix $\boldsymbol{B}$ was computed numerically as follows. Starting from the nominal lattice, the baseline optics vector $\boldsymbol{O}$ is obtained from the model corresponding to the initial PS current vector $\boldsymbol{I}$. The PS currents were then perturbed using a uniform distribution within the range $[-1\,\mathrm{A},\,1\,\mathrm{A}]$, chosen to maintain operation well within PS limits. For each perturbation $\Delta\boldsymbol{I}$, the modified lattice was evaluated using MAD-X, and the resulting change in optics $\Delta\boldsymbol{O}$ relative to the baseline lattice was computed. Subsequently, a linear regression was performed to obtain $\boldsymbol{B}$. The resulting correlation coefficients were found to be close to unity, confirming that the optics response is well approximated as linear within the chosen current range.

\begin{table*}
\centering
\caption{\label{Optics Constraints} Optics constraints used to optimize the magnet solutions.}
\begin{tabular}{@{} l l l @{}}
\toprule
\textbf{Constraint} & \textbf{Definition} & \textbf{Notes} \\
\midrule
PS Limits          & \( I^{min}_j + 5A \leq I_j \leq I^{max}_j - 5A \) & \( j \in N_{magnets} \) \\
Change in Current Strengths & \( \max(\Delta I_j) \leq 5A \)                                    & \( j \in N_{magnets} \) \\
Hysteresis                  & \( \|K^i_j\| \leq \|K^{i+1}_j\| \)                        & \( i \in N_{iterations},\ j \in N_{magnets} \) \\
Collimator at IR8          & \( \beta^i_u \leq \beta^0_u \)                            & \( i \in N_{iterations},\ u \in \{x, y\} \) \\
\bottomrule
\end{tabular}
\end{table*}

As we will demonstrate later in the paper, to move $s^*_x$ while keeping the other optical quantities constant, the currents required can be calculated from the pseudo-inverse of $\boldsymbol{B}$ from Equation~(\ref{inverse sensitivity relation}). This can be extended to moving $s^*_y$ or any other optical quantities in IR while maintaining certain constraints.

\begin{equation}\label{inverse sensitivity relation}
    \Delta \boldsymbol{I_0} = \boldsymbol{B}^{-1} \begin{bmatrix}
       \Delta s^*_x \\
       0 \\
       ... \\
       0 
       \end{bmatrix}
\end{equation}

Altering magnet strengths would result in hysteresis effects if the change in magnet strengths were to change directions for the next $\Delta s^*$ movement. To avoid hysteresis effects, a monotonicity constraint was imposed on the movement of magnet strengths, ensuring a consistent direction of values throughout the adjustment process. After each $\Delta s^*$ movement, the allowable bounds on the magnet strengths were updated to respect this constraint. Additionally, to ensure safe operating conditions, a constraint was applied to the PS currents, requiring them to remain within $\pm 5A$ from their specified limits ([$\boldsymbol{I_{min}}+5A, \boldsymbol{I_{max}} - 5A$]). When moving $s^*$, the adjustments to individual currents were also minimized in the penalty function to remain within a linear response regime and prevent abrupt transitions. We observed that the maximum $\Delta I_j$ during each movement did not exceed $5A$ after the optimization was done. These constraints are defined in Table~\ref{Optics Constraints}. Respecting these constraints was made possible using the null space of $\boldsymbol{B}$ using (\ref{inverse sensitivity relation with Null}), as Equation~(\ref{sensitivity relation}) is an under-determined system.


\begin{equation}\label{inverse sensitivity relation with Null}
    \Delta \boldsymbol{I_C} = \boldsymbol{B}^{-1} \begin{bmatrix}
       s^*_x \\
       0 \\
       ... \\
       0 
       \end{bmatrix} - \mathrm{Null}(\boldsymbol{B})\boldsymbol{C}
\end{equation}

\noindent where $\boldsymbol{C}$ is a vector of constants whose size is the number of free parameters in the system (in this case, four). These constants are chosen such that the constraints in Table~\ref{Optics Constraints} are respected. The null space accesses infinite solutions near $\boldsymbol{I_0}$ such that the optics requirements are unchanged:
\begin{equation}
 \boldsymbol{B} \Delta \boldsymbol{I_C} = \boldsymbol{B} \Delta \boldsymbol{I_0} 
\end{equation}

This optimization method also has the advantage of applying other constraints to mitigate beam losses. The beta function at collimators near IR8 were made sure to be below the original value to reduce risk of beam loss and collimator damage as shown in Table~\ref{Optics Constraints}. 
However, there is no guarantee that a solution can be found for a given optics movement depending on the initial lattice and the constraints employed, and one may have to relax the constraints or move to a nonlinear method with a well-defined inverse.


Each movement of $s^*$ produces the resulting current values, optics, and magnet strengths. The current values are inspected to confirm no current limits were violated; all other values were inspected to confirm optimization, correctness, and accuracy. We have demonstrated that solutions can be found in range of $\pm 0.5$ meters for the proton-proton lattice and are sufficient for our purpose of sPHENIX collision point optimization. The resulting magnet strengths are then stored in trim files for each movement of $s^*_x$ and sent to machine during experiment. 

\section{\label{sec: Optics Measurement Methods} Optics Measurement Methods}
\begin{figure}[htbp]
    \centering
    \includegraphics[width=.7\columnwidth]{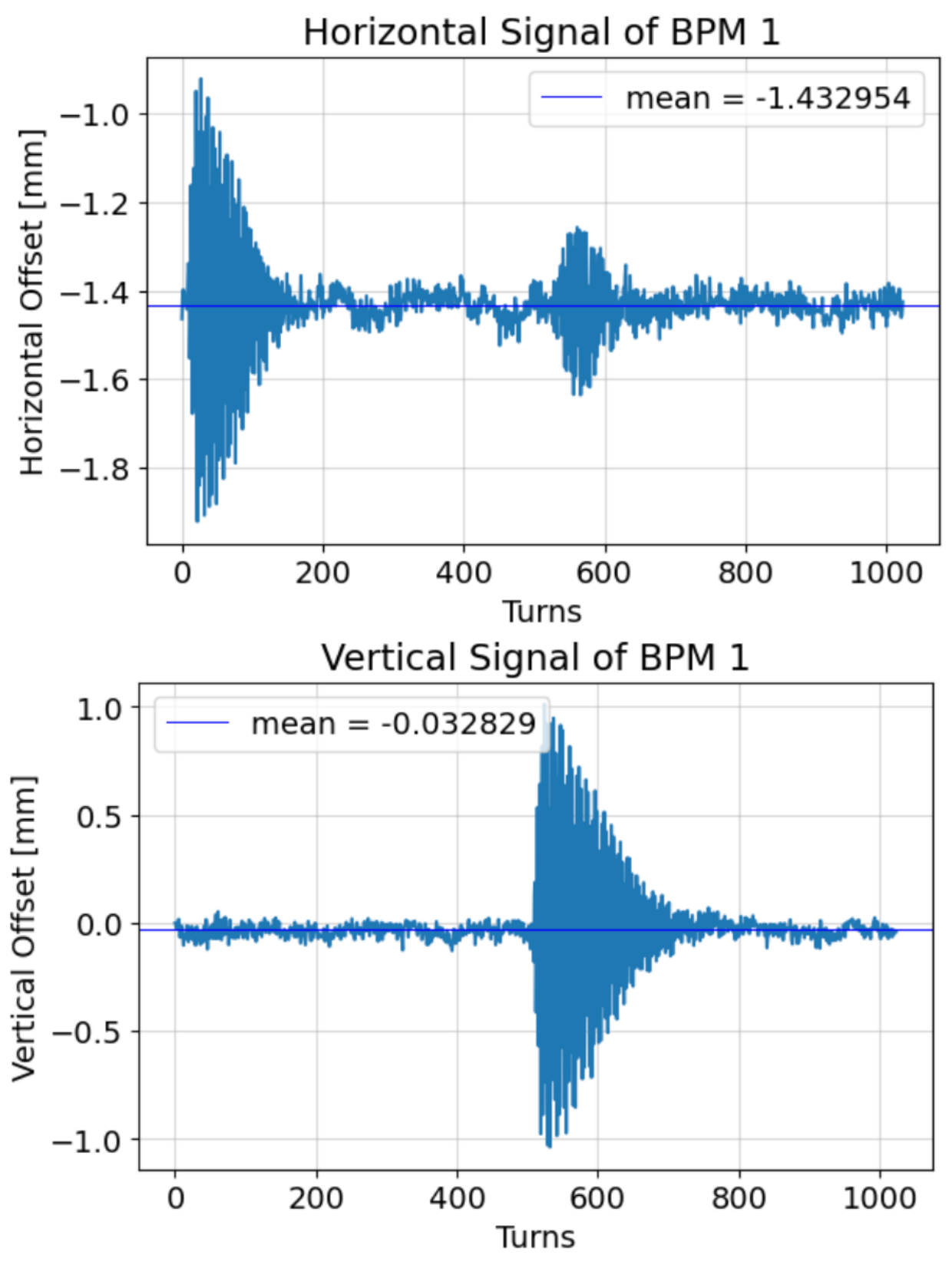}
    \caption{\label{fig:Raw_TBT_Data} 
    Horizontal and vertical TBT data from BPM upstream of IP6 for 1024 turns.}
\end{figure}

The TBT data is generated after the beam is kicked by a tune meter kicker around IR6 \cite{Liu}. The data shown in Figure~\ref{fig:Raw_TBT_Data} upstream of IR6 yields centroid positions for 1024 turns in 168 and 167 horizontal and vertical BPMs respectively. A horizontal kick is first given to the beam in the horizontal plane, then a vertical kick is applied after 500 turns. 

The TBT data is first centered to emphasize deviations from the original trajectory, allowing for clearer observation of beam responses. This is most clearly pronounced near the IRs. The data is also truncated to the first 200 turns in the horizontal plane and from turn 500-700 in the vertical plane, corresponding to the onset of full beam decoherence and coupling. At RHIC, the BPM data usually oscillates less than 500 turns due to decoherence effects \cite{Liu}. Since the RHIC AC dipoles are no longer routinely available for optics measurements, the data in this study has been analyzed with decoherence from nonlinear effects as well as chromatic detuning in mind \cite{Decoherence}. During optics measurement, the horizontal and vertical planes were treated separately while following the same analysis procedures.

BPMs exhibiting tune deviations significantly different from the nominal machine working point are classified as unreliable or "bad" BPMs. To preserve the integrity of the dataset, such BPMs are excluded from the analysis by zeroing their signals. A tune offset outside of $\pm 0.005$ was used as a criterion for exclusion.

\subsection{\label{sec: One Turn Map} One Turn Map}

\subsubsection{Ordinary Least Squares (OLS)}

The transverse phase-space coordinates can be reconstructed directly if there are two BPMs on both sides of the drift without magnetic elements.  In the lattice of the experiment, there are twelve drift spaces with BPMs on both ends, including four IRs (with detector solenoid off) and drift spaces downstream and upstream of each IR. 
In this paper, we focus on the analysis at the location of sPHENIX (IR8) as presented in Section~\ref{sec: Experimental Results}. The same procedure is applicable to the other drift regions.   

Within each drift region, the angle coordinate $u'_{12}$ can be calculated between the two endpoints $u_1$ and $u_2$ of the drift space (downstream and upstream BPMs) through Equation~(\ref{angle coordinate}):

\begin{equation}\label{angle coordinate}
    u'_{12} = \frac{u_2 - u_1}{L}
\end{equation}

\noindent where $L$ is the length of the drift space taken from \cite{RHICConfigManual}, and $u$ represents the horizontal or vertical axis. The equation for the one-turn map is of the form:

\begin{equation}\label{Next turn Eqn}
    \boldsymbol{Y} = \hat{\boldsymbol{M}}_j
    \boldsymbol{X}
\end{equation}

\begin{subequations}\label{Definitions of U}
\begin{equation}
    \boldsymbol{X} = \begin{bmatrix}
       u \\ u'
       \end{bmatrix}_j^i
\end{equation}
\begin{equation}
     \boldsymbol{Y} = 
     \begin{bmatrix}
       u \\ u'
       \end{bmatrix}_j^{i+1}
\end{equation}
\end{subequations}

\noindent $\boldsymbol{X}$ and $\boldsymbol{Y}$ contain position and angle coordinate information of the dataset for the $i^{th}$ and $(i+1)^{th}$ turn at the $j^{th}$ drift-space BPM, respectively, and

\begin{equation}\label{One Turn Map}
    \hat{\boldsymbol{M}}_j =
      \begin{bmatrix}
       \cos\phi + \alpha\sin\phi&
       \beta\sin\phi \\
       -\gamma\sin\phi &
       \cos\phi - \alpha\sin\phi 
       \end{bmatrix}_j
\end{equation}

\noindent The matrix $\hat{\boldsymbol{M}}_j$ denotes a quantity estimated from the measured data. All other quantities are assumed to be derived from measured data unless otherwise stated.

Equation~(\ref{Next turn Eqn}) represents a standard linear regression problem. This can be solved using an ordinary least squares regression (OLS) \cite{LR}:

\begin{equation}\label{OLS}
    \hat{\boldsymbol{M}}_j = (\boldsymbol{X}^T\boldsymbol{X})^{-1}\boldsymbol{X}^T\boldsymbol{Y}
\end{equation}

This was used to calculate the components of $\hat{\boldsymbol{M}}_j$ at each drift space. The linear optics ($\phi, \alpha, \beta$) at the $j^{th}$ BPM can then be computed using the elements of $\hat{\boldsymbol{M}}_j$:

\begin{subequations}\label{Optics from LR}
\begin{equation}
    \phi = \cos^{-1}\!\left( \frac{\hat{M}^{11}_j + \hat{M}^{22}_j}{2} \right)
\end{equation}
\begin{equation}
    \beta = \frac{\hat{M}^{12}_j}{\sin\phi}
\end{equation}
\begin{equation}
    \alpha = \frac{\hat{M}^{11}_j - \hat{M}^{22}_j}{2\sin\phi}
\end{equation}
\end{subequations}

Using OLS on RHIC BPM data is analogous to moment calculations of the BPM data used in \cite{Syphers, Wiedemann} since both measure linear optics from TBT data. An additional 4D parameterization of the one-turn map was also calculated to address concerns regarding coupling effects according to Figure~\ref{fig:Raw_TBT_Data} \cite{Edwards_Teng}; however, no significant coupling effects were observed regarding these measurements.

\subsubsection{Generalized Total Least Squares (GTLS)}

OLS assumes linearity in the regression model, uncorrelated and homoscedastic (constant) errors, no perfect multicollinearity among independent variables, and measurement-error-free independent variables (no errors-in-variables assumption) \cite{LR}. Linearity between $\boldsymbol{X}$ and $\boldsymbol{Y}$ is respected since they are related via drift space, and nonlinear effects are negligible compared to BPM errors.





Due to the construction of Equation~(\ref{angle coordinate}), the independent variables within $\boldsymbol{X}$ ($u_i, u'_i$) share common BPM measurements, which induces correlated measurement errors between the columns of $\boldsymbol{X}$. However, since $\boldsymbol{X}$ is still full rank and the regressors remain linearly independent, the independent variables do not introduce perfect multicollinearity. Nevertheless, the shared dependence on common BPM data induces correlated and heteroskedastic measurement errors between the reconstructed phase-space coordinates. The covariance matrix $\boldsymbol{\Sigma}$ of the independent variables' errors from Equation~(\ref{covariance}), assuming a BPM position uncertainty $\sigma_u$, explicitly shows unequal (heteroskedastic) and correlated errors between $u \text{ and } u'$.

\begin{equation}\label{covariance}
    \boldsymbol{\Sigma} = 
    \begin{bmatrix}
       \sigma_u^2 & -\frac{\sigma_u^2}{L} \\
       -\frac{\sigma_u^2}{L} & \frac{2\sigma_u^2}{L^2} 
    \end{bmatrix}
\end{equation}

Furthermore, since BPM data is used in both $\boldsymbol{X}$ and $\boldsymbol{Y}$ for OLS, this violates the assumption of error-free regressors. As a consequence, as the error in $\boldsymbol{X}$ increases, the estimated $\boldsymbol{\hat{M_j}}$ becomes increasingly biased downward, leading to an underestimation of the true $\boldsymbol{M_j}$. This is known as attenuation bias \cite{LR}:

\begin{equation}\label{attenuation bias}
    \boldsymbol{\hat{M}_j} = \boldsymbol{M}_j\frac{\sigma_r^2}{\sigma_{r}^2 + \sigma_{u}^2}
\end{equation}

Equation~(\ref{attenuation bias}) demonstrates a single-variable expression for attenuation bias. Since our situation is multivariate, this bias depends more generally on $\boldsymbol{\Sigma}$. OLS is expected to perform well when the errors in $\boldsymbol{X}$ ($\sigma_u$) are negligible compared to the intrinsic spread of true regressors $\sigma_r$ \cite{LR}. Since this is not the case as shown in Figure~\ref{fig:LR_vs_noise}, an error-in-variables treatment is required.

A generalized total least squares (GTLS) formulation could be utilized to address the violations of error-in-variables, correlated and heteroskedastic measurement errors in $\boldsymbol{X} \text{ and } \boldsymbol{Y}$. The algorithm of GTLS is explained in more detail in \cite{GTLS}, but a high level overview will be presented here, with each step explaining how OLS is improved.

First, $\boldsymbol{X} \text{ and } \boldsymbol{Y}$ are whitened according to Equation~(\ref{covariance}). The covariance matrix for $\boldsymbol{X} \text{ and } \boldsymbol{Y}$ is assumed to be the same since they were constructed from the same BPM data. Specifically, a Cholesky decomposition can be applied to $\boldsymbol{\Sigma}$ ($\boldsymbol{\Sigma} = \boldsymbol{L}\boldsymbol{L}^{T}$), and each $\boldsymbol{L}$ can be used to whiten $\boldsymbol{X} \text{ and } \boldsymbol{Y}$ ($\boldsymbol{X}_w = \boldsymbol{X}\boldsymbol{L}^{-T} \text{ and } \boldsymbol{Y}_w = \boldsymbol{Y}\boldsymbol{L}^{-T}$). The whitened variables $\boldsymbol{X}_w \text{ and } \boldsymbol{Y}_w$ are such that their covariance matrices are the identity matrix, which makes their errors constant and uncorrelated in this whitened space.

Instead of regular OLS, a total least squares (TLS) procedure is applied to $\boldsymbol{X}_w \text{ and } \boldsymbol{Y}_w$ to find a measurement of $\hat{\boldsymbol{M}}$. 
Under conditions guaranteeing a unique solution (see \cite{TLS}), TLS can be obtained from the singular value decomposition (SVD) of the concatenated data matrix,

\begin{equation}\label{TLS SVD}
    \begin{bmatrix}
        \boldsymbol{X}_w & \boldsymbol{Y}_w
    \end{bmatrix} = 
    \begin{bmatrix}
        \boldsymbol{U}_X & \boldsymbol{U}_Y
    \end{bmatrix}
    \begin{bmatrix}
        \boldsymbol{\sigma}_x & 0 \\ 
        0 & \boldsymbol{\sigma}_Y
    \end{bmatrix}
    \begin{bmatrix}
        \boldsymbol{V}_{XX} & \boldsymbol{V}_{XY} \\ 
        \boldsymbol{V}_{YX} & \boldsymbol{V}_{YY}
    \end{bmatrix}^*
\end{equation}


\noindent where the right-singular-vector matrix is partitioned conformably with $\boldsymbol{X}_w$ and $\boldsymbol{Y}_w$. The whitened one-turn map at the $j$th BPM, $\hat{\boldsymbol{M}}_{j,w}$, is then

\begin{equation}\label{TLS Sol}
    \hat{\boldsymbol{M}}_{j, w} = -\boldsymbol{V}_{XY}\boldsymbol{V}_{YY}^{-1}
\end{equation}

The TLS algorithm aims to reduce the noise in both $\boldsymbol{X}_w$ and $\boldsymbol{Y}_w$ by selecting dominant, orthogonal modes that contribute to the one-turn map while discarding low variance modes corresponding to BPM errors. This SVD approach is how TLS deals with errors-in-variables. The resulting map is transformed back to the original coordinates using $\boldsymbol{L}$ ($\hat{\boldsymbol{M}}_{j}$ = $\boldsymbol{L}^{-T}\hat{\boldsymbol{M}}_{j, w}\boldsymbol{L}^{T}$). The improvement in the $\beta$ estimate obtained by using GTLS instead of OLS with respect to BPM noise is shown from Figure~\ref{fig:LR_vs_noise}.

\subsection{\label{sec: Curve Fit} Curve Fit}

In this paper, we refer to least-squares fitting TBT BPM data as the \emph{curve-fitting} (CF) method. CF is a standard implementation of the ``beta-from-amplitude'' technique, where the beta function is inferred from the fitted betatron oscillation amplitude at each BPM. Since the RHIC BPM signals exhibit significant decoherence, the fitting model must account for this effect in order to faithfully represent the measured beam dynamics.

Following the decoherence model in \cite{Decoherence}, we employ a simplified form that retains the essential structure of closed-orbit oscillations while reducing the number of free parameters for nonlinear least-squares fitting. In particular, we include decoherence due to chromatic effects only. 
The BPM signal over $n$ turns is modeled with an amplitude $a$, a decoherence coefficient $b$, the machine tune $\nu$, and the phase offset $\phi$. 

\begin{equation}\label{Closed orbit fit}
    x_{co}(n) = a\exp\left(\frac{-2\pi n^2}{b^2}\right)\cos(2\pi\nu n + \phi)
\end{equation}

This nonlinear least-squares fit is applied to all BPMs excluding the unreliable BPMs. The amplitudes of each BPM are then collected into a vector $\boldsymbol{A}$ and the beta function is proportional to the square of the amplitudes. A ratio (similar to the action \cite{coupling}) is then calculated from the amplitudes and model beta function from arc sections and multiplied to the square of the amplitudes from Equation~(\ref{factor}) to get the measured beta values from CF ($\boldsymbol{\beta}_{meas}^{CF}$) over the whole ring:

\begin{equation}\label{factor}
    \boldsymbol{\beta}_{meas}^{CF} = \boldsymbol{A}^2
    \frac{\langle \boldsymbol{\beta}_{model, arc} \rangle}
    {\langle \boldsymbol{A}^2_{arc}\rangle}
\end{equation}

\noindent where $\langle \dots \rangle$ is taken to be the expected value. The arc section values were chosen due to the smaller values of the beta function in those regions generally yielding less errors than those in the IRs \cite{ICARHIC}. An error analysis between LR, CF, and HA methods is given in Section~\ref{sec: Errors of Optics Measurement Methods}.

\subsection{\label{sec: Linear Optics at IP} Linear Optics at IP}
The linear optics parameters $\beta^*$ and $s^*$ at the IP were extracted using the $\beta$-function measured at BPMs located at the IRs. Near a waist, the $\beta$-function follows the standard quadratic form \cite{Wiedemann}.

\begin{equation}\label{beta at IR}
    \beta_{IR}(s) = \beta^* + \frac{(s - s^*)^2}{\beta^*}
\end{equation}

Two approaches were used to determine $\beta^*$ and $s^*$: A nonlinear least squares fit using $\beta_{meas} \text{ and } s$ at both ends of the IR and a middle-layer calculation of the twiss parameter $\alpha$, derived from Equation~(\ref{beta at IR}).

\begin{subequations}\label{alpha at IR}
    \begin{align}
        \beta^* = \frac{\beta_1}{1 + \alpha_1^2}
    \end{align}
    \begin{align}
        s^* = s_1 + \beta^*\alpha_1
    \end{align}
\end{subequations}
 
When the $\beta$ value and longitudinal location of an upstream and downstream BPM are used ($(\beta_1, s_1)\text{ and } (\beta_2, s_2)$ respectively), there are two mathematical solutions; one that is physical for IRs and generally near $\beta_{model}$ and one far from $\beta_{model}$. Because of this, selecting the physically consistent branch requires additional information. We required $\beta^*_{meas}$ to be close to an initial guess $\beta^*_{model}$ for continuity with the nominal optics, introducing a weak model dependence. On the other hand, if $\alpha$ is reconstructed from the linear optics (e.g. via GTLS), this method is model-independent with respect to the lattice optics used for branch selection.


A detailed error analysis of Equation~(\ref{beta at IR}) is presented in \ref{sec: Errors of Optics Measurement Methods}. Using Equation~(\ref{alpha at IR}) increases the error of $\beta^* \text{ and } s^*$ for every optics measurement method. This error amplification was found to be more pronounced in CF and HA than GTLS, as shown in Figure~\ref{fig:stdv_star_alpha}. Therefore when calculating $\beta^*$ and $s^*$, GTLS uses Equation~(\ref{alpha at IR}) to keep the method independent of a model while CF and HA utilize Equation~(\ref{beta at IR}) to avoid significant error amplification.



\section{\label{sec: Experimental Results} Experimental Results}

The beam for this study consisted of 12 proton bunches in the blue ring at $100~\mathrm{GeV}$, with $1\times10^{11}$ particles per bunch. The TBT datasets used in this study were acquired during RHIC fill 34486 in 2024. After the beams were in store, initial TBT measurements were taken. Since the insertion quadrupoles at IP8 have been unaltered, the data taken corresponds to $\Delta s^* = 0$. A measurement similar to that for $\beta^*$ beat and $s^*$ difference using R-OP in Table~\ref{R_IP8_Metrics} was also done for CF, HA, and GTLS. The same six initial TBT measurements corresponding to $\Delta s^* = 0$ were analyzed and the same metrics were taken. These results are shown in Table~\ref{LOMM_IP8_Metrics_0}.

\begin{table*}
\caption{\label{LOMM_IP8_Metrics_0} The mean absolute value of $\beta^*$ beat, mean $s^*$ difference, and  sample standard deviations for both at IR8 were calculated over thes same six TBT datasets, this time for each optics measurement method. The mean absolute value was taken via Equation~(\ref{mean abs val}), where $N$ is the number of TBT datasets. Error bars are calculated from Section~\ref{sec: Error Analysis}}.
\centering
\begin{tabular}{c 
            @{\hspace{12pt}} r@{\,\( \pm \)\,}l 
            @{\hspace{24pt}} r@{\,\( \pm \)\,}l 
            @{\hspace{24pt}} r@{\,\( \pm \)\,}l 
            @{\hspace{24pt}} r@{\,\( \pm \)\,}l}
\toprule
Method & 
\multicolumn{2}{c}{$\bar\beta^*_{x, avg}$ Beat [\%]} & \multicolumn{2}{c}{$\bar{s}^*_x$ Difference [m]} & 
\multicolumn{2}{c}{$\bar{\beta}^*_{y, avg}$ Beat [\%]} & \multicolumn{2}{c}{$\bar{s}^*_y$ Difference [m]} \\
\midrule



R-OP  & 18    & 7    & 0.0  & 0.4  & 7    & 5    & 0.1  & 0.2  \\
GTLS  & 8.0   & 9.5  & 0.09 & 0.1  & 5.7  & 3    & -0.05 & 0.03 \\
CF    & 6.6   & 4.2  & 0.12 & 0.1  & 1.1  & 1.9  & -0.07 & 0.04 \\
HA    & 3.8   & 4.2  & 0.14 & 0.1  & 1.4  & 2.0  & -0.06 & 0.03 \\

\bottomrule
\end{tabular}
\end{table*}

Overall, CF, HA, and GTLS outperform R-OP in measuring these quantities. On average, the $\beta^*$ beat is below $10 \%$ and while $s^*_x$ is shown to measure slightly off the expected value, the variation in $s^*$ is around or below $10$ cm for all three methods. The error in $\beta^*$ beat for GTLS is comparable in both planes to R-OP; however, GTLS measures a smaller $\beta^*$ beat compared to R-OP. Furthermore, the high variance at IR8 is consistent with the error analysis at high-$\beta$ regions; at low-$\beta$ regions, GTLS has a lower $\beta^*$ beat error as shown in Figure~\ref{fig:stdv_star} and Figure~\ref{fig:stdv_star_alpha}. The sample standard deviation measurements also align with the Monte Carlo simulation and analytical error propagation studies when $s^*$ is not moved in \ref{sec: Errors of Optics Measurement Methods}.

When moving $s^*$, the magnet strengths were adjusted in the control room by $\Delta s^*_x = 0.1$m according to the sensitivity matrix trim file corresponding to the appropriate $\Delta s^*_x$ movement. After the movement, two TBT measurements were taken. The beam loss rate, tune, and beta around collimators are then monitored after every $\Delta s^*_x$ movement to ensure no significant beam loss. The magnet strengths are also monitored to ensure correctness and to confirm that magnet strengths have been altered. This procedure was repeated afterward for $\Delta s^*_x = 0.3\text{m}, 0.5\text{m}$ and again for backwards movement: $\Delta s^*_x = -0.1\text{m}, -0.3\text{m}, -0.5\text{m}$.

\begin{figure*}[htbp]
    \centering
    \includegraphics*[width=2\columnwidth]{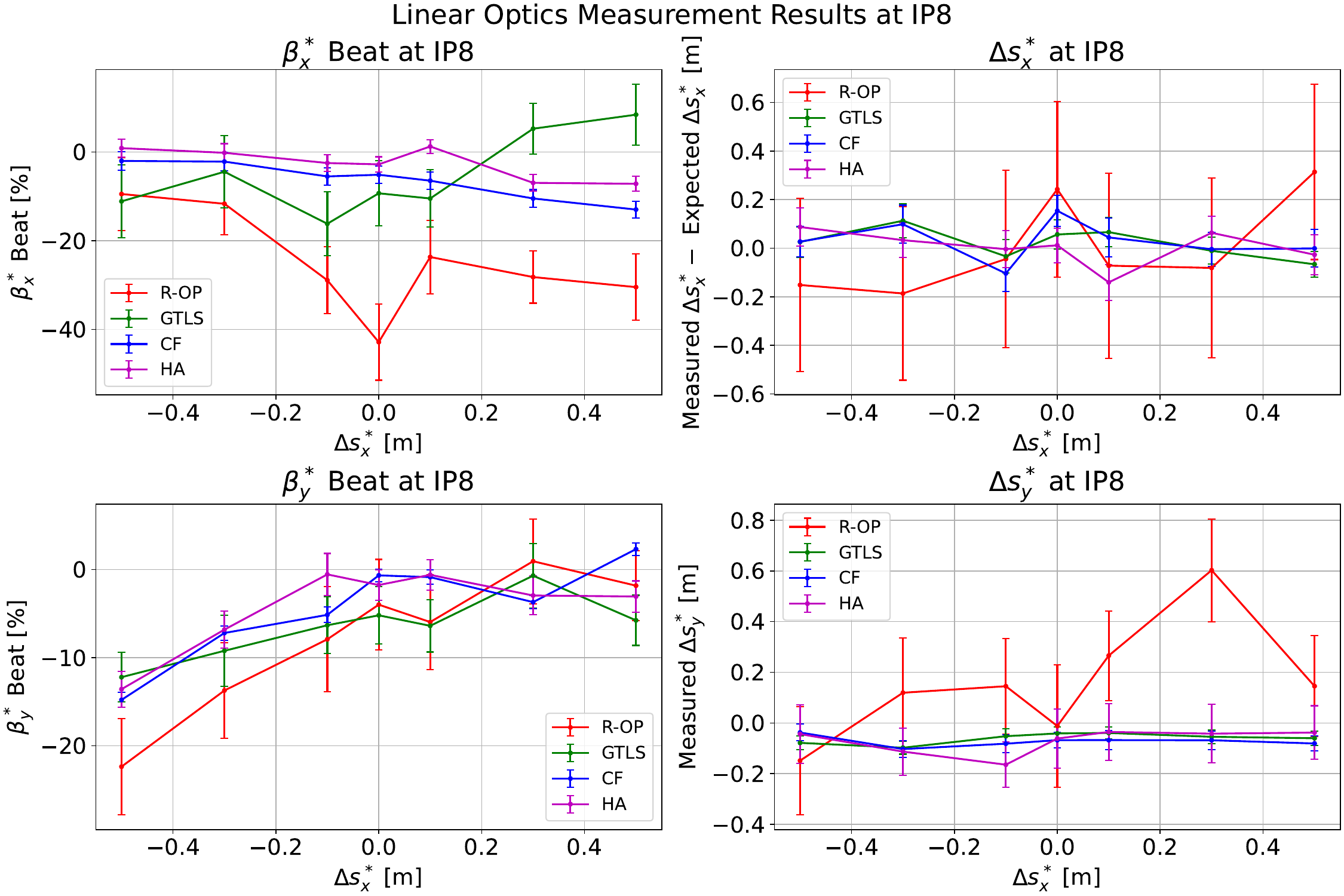}
    \caption{\label{fig:LOMM_IP8_err} 
    The average between two measurements for each $\Delta s^*_x$ movement were taken and plotted. Measurements of $\beta^*$ beat and $s^*$ difference at IP8 were calculated using R-OP, GTLS, CF, and HA. The expected value of $\Delta s^*_x$ was subtracted out of the measured values for consistent structure with the other graphs (Measured $\Delta s^*_x$ is expected to match the expected $\Delta s^*$ movement; The expected $\Delta \beta^*_x, \Delta \beta^*_y, \Delta s^*_y$ movements are $0$). A BPM error level of 30 $\mathrm{\mu m}$ was assumed for IP8, and error bars are calculated from Section~\ref{sec: Error Analysis}.}
\end{figure*}

After the TBT data had been gathered, the linear optics around IRs were then measured through the RHIC optics program (R-OP) as well as GTLS, CF, and HA. Upon observing the $s^*$ measurements at IP8, all methods demonstrate on average the correct general trend of appropriate $s^*$ movement. This can be seen in Figure~\ref{fig:LOMM_IP8_err} for $\Delta s^*$ movement in both planes; after subtracting the expected motion, the residual $s^*$ differences cluster around 0. This validates the methods mentioned in Section~\ref{sec: Optics tuning}. However, each method shows distinct results in terms of beta beat and $s^*$ difference.

At each $\Delta s^*$ movement, a comparison between all of the optics measurement methods are made for $\beta^*$ and $s^*$ around the IRs. Figure~\ref{fig:LOMM_IP8_err} displays a plot of the experimental calculations for $\beta^*$ beat and $s^*$ difference in both planes at IP8, measured by the optics measurement methods with error bars calculated from \ref{sec: Errors of Optics Measurement Methods}. A BPM noise level of 30 $\mathrm{\mu m}$ was assumed, consistent with previous RHIC studies \cite{ICARHIC}, where non-arc BPMs have slightly more error on average than arc BPMs at RHIC. The error bars for R-OP were calculated similarly to Table~\ref{R_IP8_Metrics}. From the figure, GTLS, CF, and HA all overall measure a smaller $\beta^*$ beat and $s^*$ difference than R-OP. Furthermore, their measurements show smaller residual offsets and variance relative to R-OP as well as consistency with one another, indicating improved agreement with the expected optics.


This is further quantitatively shown in Table~\ref{LOMM_IP8_Metrics}, where the $\beta^* \text{ beat and } s^*$ difference values from Figure~\ref{fig:LOMM_IP8_err} were averaged over $\Delta s^*_x$ movements for each optics measurement method according to Equation~(\ref{mean abs val}), similar to the results in Figure~\ref{fig:RHIC_beat} and Table~\ref{R_IP8_Metrics}. GTLS, CF, and HA show improved agreement with the expected optics of star measurements than R-OP. The $\sigma_{\beta^*}$ of GTLS and R-OP are higher than CF and HA (although GTLS is slightly lower). This was also seen in Table~\ref{LOMM_IP8_Metrics_0}; $\sigma_{\beta^*}$ of GTLS decreased, showing that the error of $\beta^*$ beat of GTLS is at least comparable to R-OP. However, $s^*$ measurements done by GTLS show comparable accuracy and robustness to CF and HA. The comparable results of both Table~\ref{LOMM_IP8_Metrics_0} and Table~\ref{LOMM_IP8_Metrics} also demonstrate the robustness of the ORM method, as average measurements while moving $s^*$ yield similar results to those when $s^*$ is not shifted.

\begin{table*}
\caption{\label{LOMM_IP8_Metrics} The mean absolute value of $\beta^*$ beat, mean $s^*$ difference, and  sample standard deviations for both at IR8 were calculated over $\Delta s^*_x$ movements from Figure~\ref{fig:LOMM_IP8_err} for each optics measurement method. The mean absolute value was taken via Equation~(\ref{mean abs val}), where N here is the number of $\Delta s^*_x$ movements. Error bars are calculated from Section~\ref{sec: Error Analysis}.}
\centering
\begin{tabular}{c 
            @{\hspace{12pt}} r@{\,\( \pm \)\,}l 
            @{\hspace{24pt}} r@{\,\( \pm \)\,}l 
            @{\hspace{24pt}} r@{\,\( \pm \)\,}l 
            @{\hspace{24pt}} r@{\,\( \pm \)\,}l}
\toprule
Method & 
\multicolumn{2}{c}{$\bar\beta^*_{x, avg}$ Beat [\%]} & \multicolumn{2}{c}{$\bar{s}^*_x$ Difference [m]} & 
\multicolumn{2}{c}{$\bar{\beta}^*_{y, avg}$ Beat [\%]} & \multicolumn{2}{c}{$\bar{s}^*_y$ Difference [m]} \\
\midrule

R-OP & 25 & 9 & 0.2 & 0.3 & 8 & 5 & 0.2 & 0.2 \\
GTLS & 9 & 7 & 0.1 & 0.1 & 7 & 3 & 0.10 & 0.03 \\
CF   & 6 & 4 & 0.1 & 0.1 & 5 & 2 & 0.10 & 0.03 \\
HA   & 3 & 4 & 0.0 & 0.1 & 4 & 3 & 0.1 & 0.1 \\

\bottomrule
\end{tabular}
\end{table*}

When comparing GTLS with CF and HA, CF and HA seem to agree more with the model than GTLS in terms of beta beat. This behavior is expected since CF and HA are model-dependent methods; the explicit use of the model in the reconstruction process constrains the solution to naturally be closer in agreement with the design optics. The errors measured are in accordance with \cite{PhysRevAccelBeams.23.042801}, and errors in $\beta^*$ beat can be amplified by gain errors as shown in Figure~\ref{fig:stdv_star_gain}. On the other hand, GTLS gives $\beta^*$ beat and $s^*$ difference measurements that agree more closely with the model than R-OP, though not as closely as CF and HA. Gain errors are also shown to not affect errors in $\beta^*$ beat as much in GTLS as in CF and HA. However, $s^*$ measurements are affected regardless of method. Since measuring absolute accuracy of the optics is not possible, GTLS, CF, and HA all present possible values for IP8 since they are consistent with each other.


Given this evaluation method, $s^*_x$ and $s^*_y$ differences show consistent measurements around the model and each measurement method besides R-OP. According to Table~\ref{LOMM_IP8_Metrics}, the error of $s^*$ difference is shown to be approximately within $\pm 10$ cm in both the horizontal and vertical directions for GTLS, CF, and HA. However, the $\beta^*_y$ beat shows that as $s^*_x$ is moved in the negative direction, the $\beta^*_y$ beat increases substantially for all methods, as seen in Figure~\ref{fig:LOMM_IP8_err}. This shows that improvements can be made in optics control, as nonlinear magnetic effects may introduce disparities between machine and model. Nevertheless, although $\boldsymbol{B}$ is not perfect, it is able to move the optics within the observed $\pm 10$ cm measurement variation between $s^*_x$ movements.

The errors obtained in $s^*$ difference from GTLS, CF, and HA show that error within $\pm 10$ cm is possible for collisions at sPHENIX vertex. For average $\beta^*$ beat in IP8, GTLS, CF, and HA is shown to predict within $10\%$. This shows a smaller $\beta^*$ beat measurement between GTLS, CF, and HA methods than R-OP and confirms the stability of beta at IP8 when moving $s^*_x$.

\section{\label{sec: Conclusion} Conclusion}

In this paper, we report the recent effort of tuning and measuring the location of the beta waist for luminosity optimization at RHIC.

The sensitivity matrix $\boldsymbol{B}$ method has been demonstrated to consistently move the magnets such that $s^*_x$ can be changed appropriately while limiting perturbations to the other constrained optics parameters as well as obeying certain machine constraints. This was verified with all linear optics methods used, as the measurements of $s^*_x$ moved appropriately while the other optics throughout the ring remained relatively unchanged. This method to move $s^*_x$ can be extended to moving $s^*$ in both transverse directions while taking care of the similar constraints as before.



For linear optics measurements, we have compared various model-independent (GTLS) and model-dependent methods (CF, HA) with R-OP (the current method used in RHIC operation) as a reference. GTLS adheres to the assumptions that OLS requires by reducing errors in variables and making errors constant and uncorrelated. The dependent and independent phase space coordinates when using OLS on a one-turn-map contain BPM errors. GTLS mitigates this effect by using covariance weighting and SVD techniques to reduce attenuation bias. Furthermore, the angle coordinates are constructed using BPM measurements, leading to non-constant and correlated errors. This is mitigated by incorporating a covariance matrix on the reconstructed phase-space coordinates to whiten the data prior to regression.

As a model-independent approach, GTLS avoids reliance on an ideal lattice and therefore provides a more robust optics measurement in the presence of model discrepancies. However, GTLS yields larger errors when measuring $\beta^*$ at high beta regions such as IP8 when compared to model-dependent methods such as CF and HA. In contrast, GTLS measures smaller errors at lower beta regions due to the different correlation coefficients calculated from each method. Furthermore, since CF and HA rely on a model, they are more sensitive to systematic effects and modeling inaccuracies. Regardless of the method, GTLS, CF, and HA yield mutually consistent measurements and reduce the average $\beta^*$ beat at IP8 to within 10\% and $s^*$ difference to within $10$ cm, representing a significant improvement over the R-OP measurement method. Analysis and cross-validation of additional linear optics measurement techniques and error analysis such as MIA\cite{MIA}, N-BPM methods\cite{NBPM2}, K-modulation techniques\cite{Carlier2017BetaStar}, as well as additional model-independent techniques\cite{Syphers} would strengthen confidence in measurements in $\beta^*$ and $s^*$ at IPs.




We have demonstrated a reliable set of methods to move and validate the waist of beta function at IP at RHIC, establishing confidence in the reliability and reproducibility of $s^*$ movement. An experiment using Bayesian Optimization (BO) to locate the optimal collision region within the sPHENIX detector has also been recently conducted. Because the precision of $s^*$ movement directly impacts the accuracy of the $s^*$ request from BO to RHIC, this work is an essential step toward improving luminosity by moving $s^*$ closer to zero in an experimental setting. These results will be reported in a separate publication.

Beyond RHIC, the methods presented here are broadly applicable to other circular accelerator and collider facilities. The Electron-Ion Collider project may benefit from these techniques due to its large sensitivity of luminosity on the position of small vertical beta waist, which is approximately 5 cm. GTLS-based reconstruction of the one-turn map would be especially helpful as it is shown to provide robust measurements in low beta regions. The techniques for linear optics movement and measurement presented in this paper form an accurate and robust framework to control and measure the linear optics near the interaction region of circular colliders.


\section*{Acknowledgements}
This work is supported by DOE Office of Nuclear Physics under award number DE-SC0023518. This work is also supported by Brookhaven Science Associates, LLC under Contract No. DE-SC0012704 and DE-AC02-06CH11357 with the U.S. Department of Energy.
\appendix

\section{\label{sec: Error Analysis} Error Analysis}
An error analysis study was done to measure the difference between OLS and GTLS as well as to measure the robustness of the optics measurement methods presented. A beam tracking simulation using Xsuite was done using the RHIC lattice \cite{Xsuite}. Two proton beams were simulated to match the amplitude and decoherence of RHIC BPM data for 200 turns. These were then tracked for 168 horizontal and 167 vertical BPMs. BPM errors were assumed to be gaussian and propagated through Monte Carlo simulations, with the noise added in as:

\begin{equation}
    u_{meas} = u_{true} + \sigma_{u}
\end{equation}

\noindent where $\sigma_{u}$ is the BPM noise. The error analysis procedure was applied to both planes; the results for only the horizontal plane is presented in this and all subsequent sections for clarity.

\subsubsection{\label{sec: OLS vs GTLS} OLS vs GTLS}
\begin{figure}[htbp]
    \centering
    \includegraphics[width=.9\columnwidth]{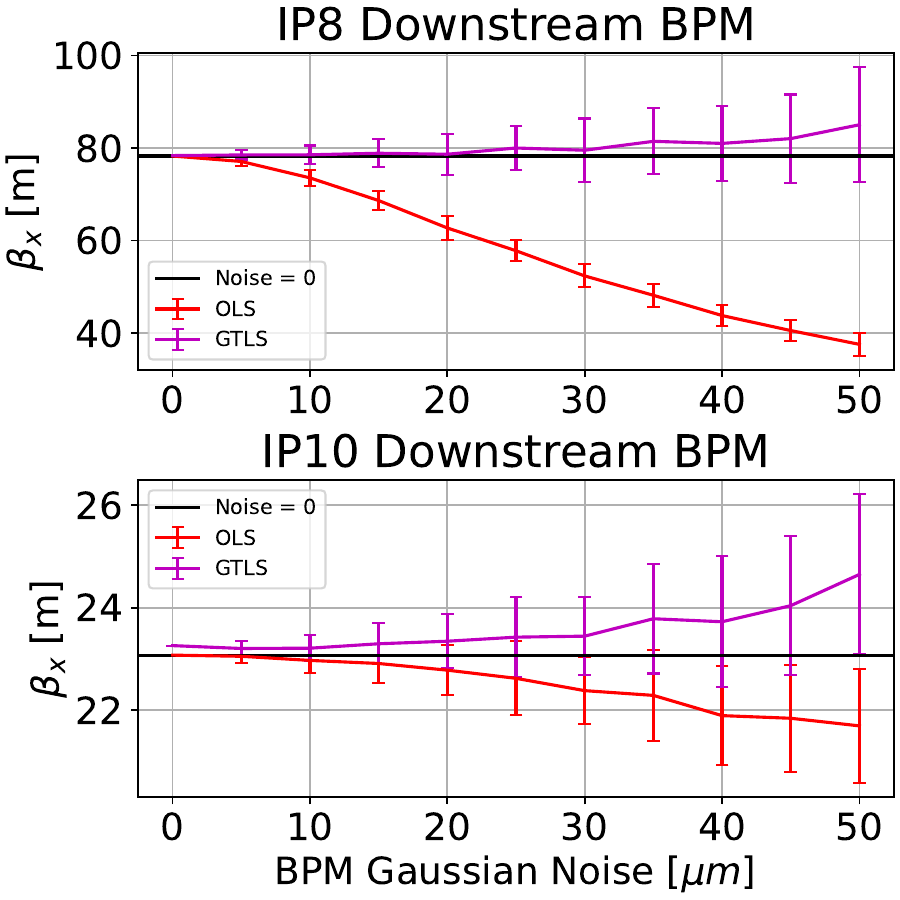}
    \caption{\label{fig:LR_vs_noise} 
    Simulated beta measurements from OLS (red) and  GTLS (purple) vs BPM error level from 0 to 50$\mathrm{\mu m}$. The solid black lines indicate measurement with no noise at a BPM in a high beta region (top) and low beta region (bottom). Error bars were calculated using a Monte Carlo simulation of 500 iterations for each noise level.}
\end{figure}

A Monte Carlo simulation of 500 iterations was ran to compare beta measurements from OLS and GTLS against BPM noise, shown in Figure~\ref{fig:LR_vs_noise}. A baseline where no noise is applied is shown as a black line. The results demonstrate that as the BPM noise level increases, beta measurements from GTLS stay close to a noiseless measurement compared to OLS in high beta (IP6, IP8) and and low beta (IP10, IP12) regions. The larger error bar sizes compared to OLS is due to the correlation in measurements between two BPMs, and is further amplified from the SVD step when implementing GTLS.


\begin{figure*}[htbp]
    \centering
    \includegraphics[width=1.5\columnwidth]{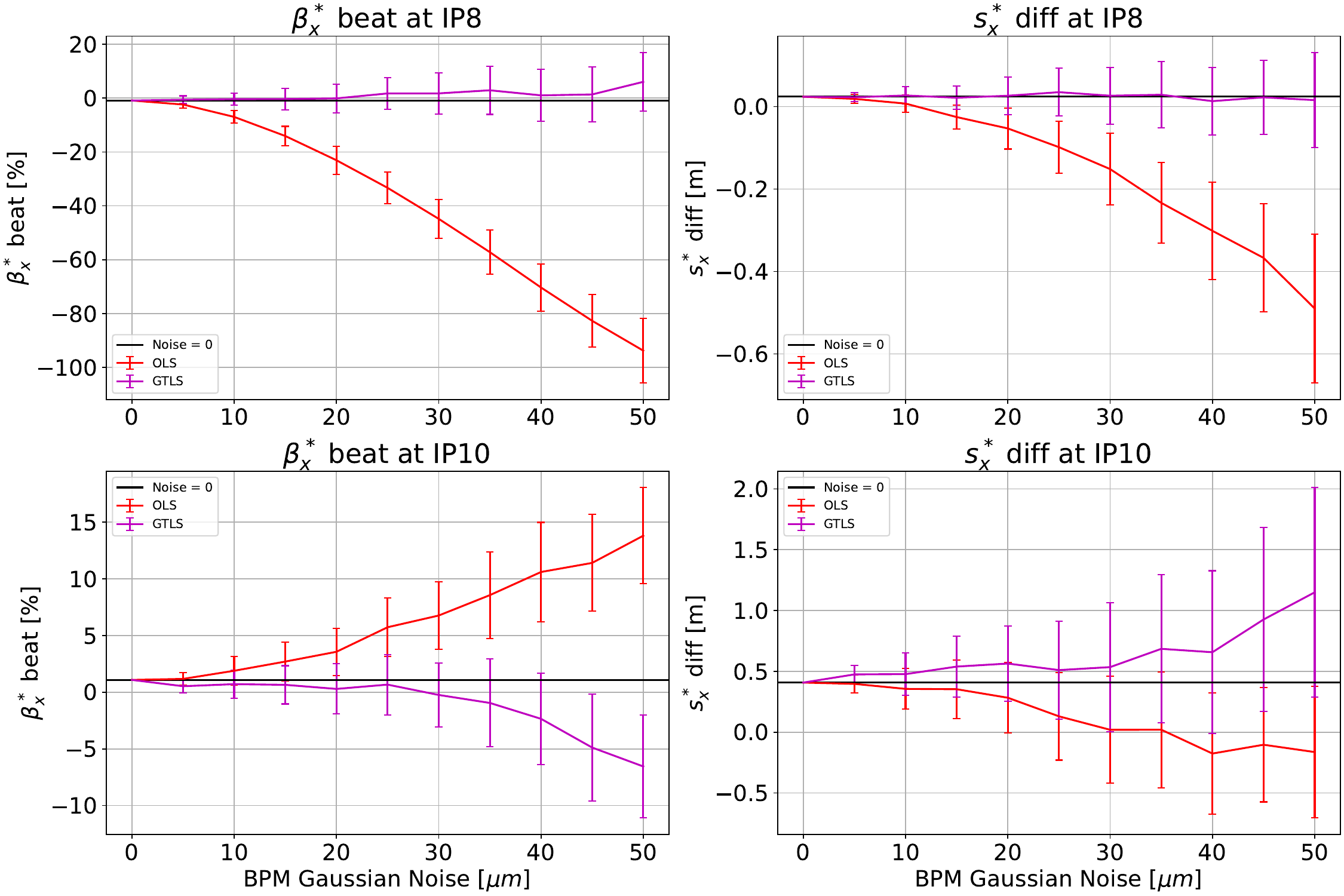}
    \caption{\label{fig:LR_vs_noise_star} 
    Simulated $\beta^*$ beat and $s^*$ difference measurements from OLS (red) and  GTLS (purple) vs BPM error level from 0 to 50$\mathrm{\mu m}$. The solid black lines indicate measurement with no noise at a BPM in a high beta region (top) and low beta region (bottom). Error bars were calculated using a Monte Carlo simulation of 500 iterations for each noise level.}
\end{figure*}

The $\beta^*$ beat and $s^*$ difference is also calculated from beta measurements using OLS and GTLS shown in Figure~\ref{fig:LR_vs_noise_star}. At $\sigma_u \leq 30 \mathrm{\mu m}$, GTLS measures a smaller $\beta^*$ beat and $s^*$ difference than OLS. This further shows the necessity to use GTLS to measure IP values.

\subsubsection{\label{sec: Errors of Optics Measurement Methods} Errors of Optics Measurement Methods}

The CF method and a harmonic analysis (HA) method from \cite{Castro} using a Fourier decomposition of BPM data were used to compare to the measurement results of GTLS. The same particle tracking was used, and errors are once again added to the BPM signals. These errors are propagated into the $\beta$ measurement as well as $\beta^*$ and $s^*$. This was all done using 100 Monte Carlo iterations for each noise level.

\begin{figure*}[htbp]
    \centering
    \includegraphics[width=1.5\columnwidth]{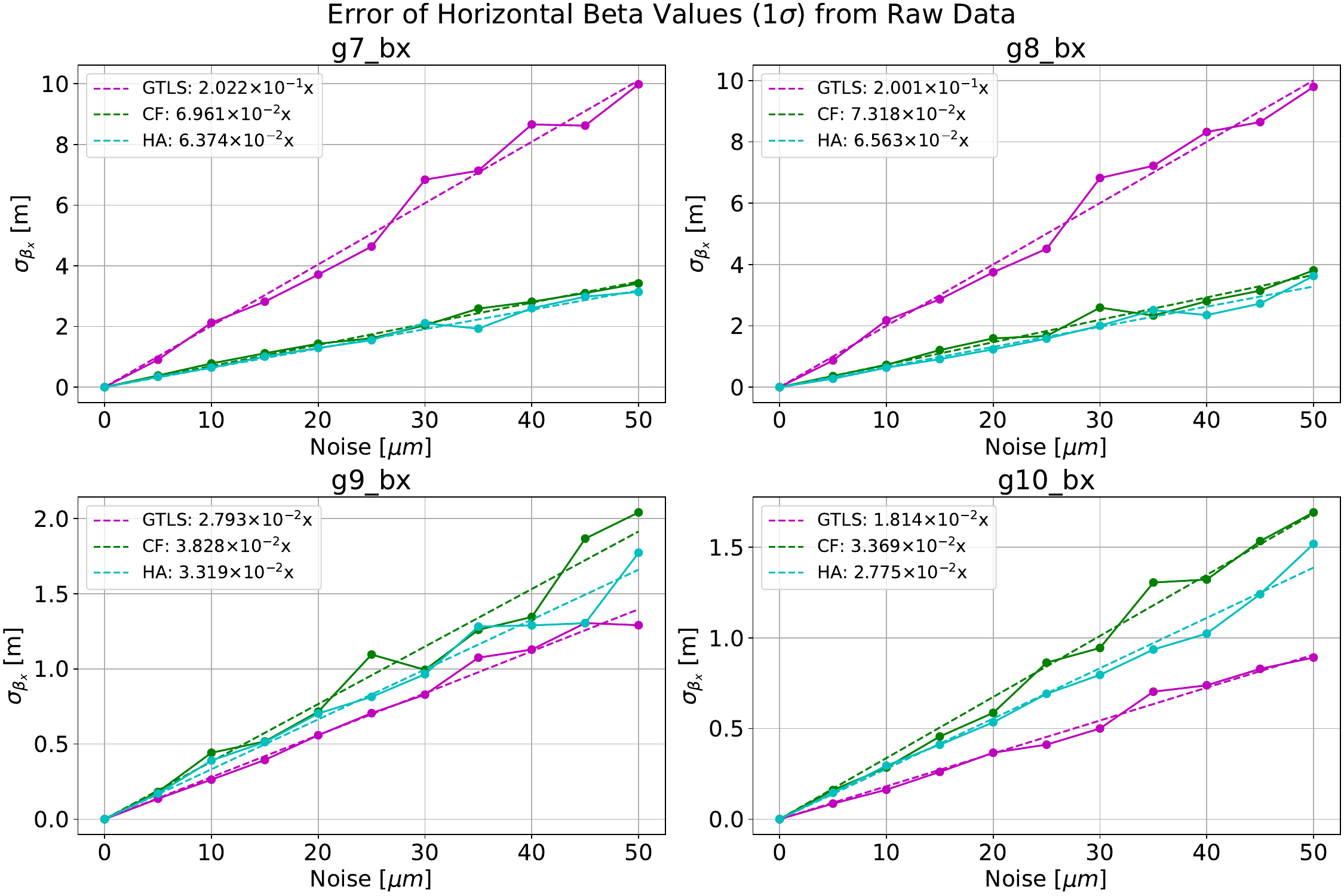}
    \caption{\label{fig:stdv_beta} 
    Error trends of GTLS (magenta), CF (green), and HA (cyan) vs BPM error level from 0 to 50$\mathrm{\mu m}$ for the measured $\beta$ value of the BPM downstream (top) and upstream (bottom) for IP8 (top) and IP10 (bottom). The points represent the standard deviation values used for the linear regression calculation of the error trends, where $x$ in the legend is the noise level.}
\end{figure*}

A plot of the standard deviations for the upstream and downstream BPMs at a high (IP8) and low beta (IP10) region is shown in Figure~\ref{fig:stdv_beta}. The trend of these standard deviations are expected to be linear, and a linear regression was used to calculate the error trends. CF and HA show a smaller standard deviation than LR methods at high beta regions and vice versa at low beta regions. From this, it seems the intermediate step of calculating the one-turn map may contribute to the increase in error bars if the beta region is high.

\begin{figure*}[htbp]
    \centering
    \includegraphics[width=1.5\columnwidth]{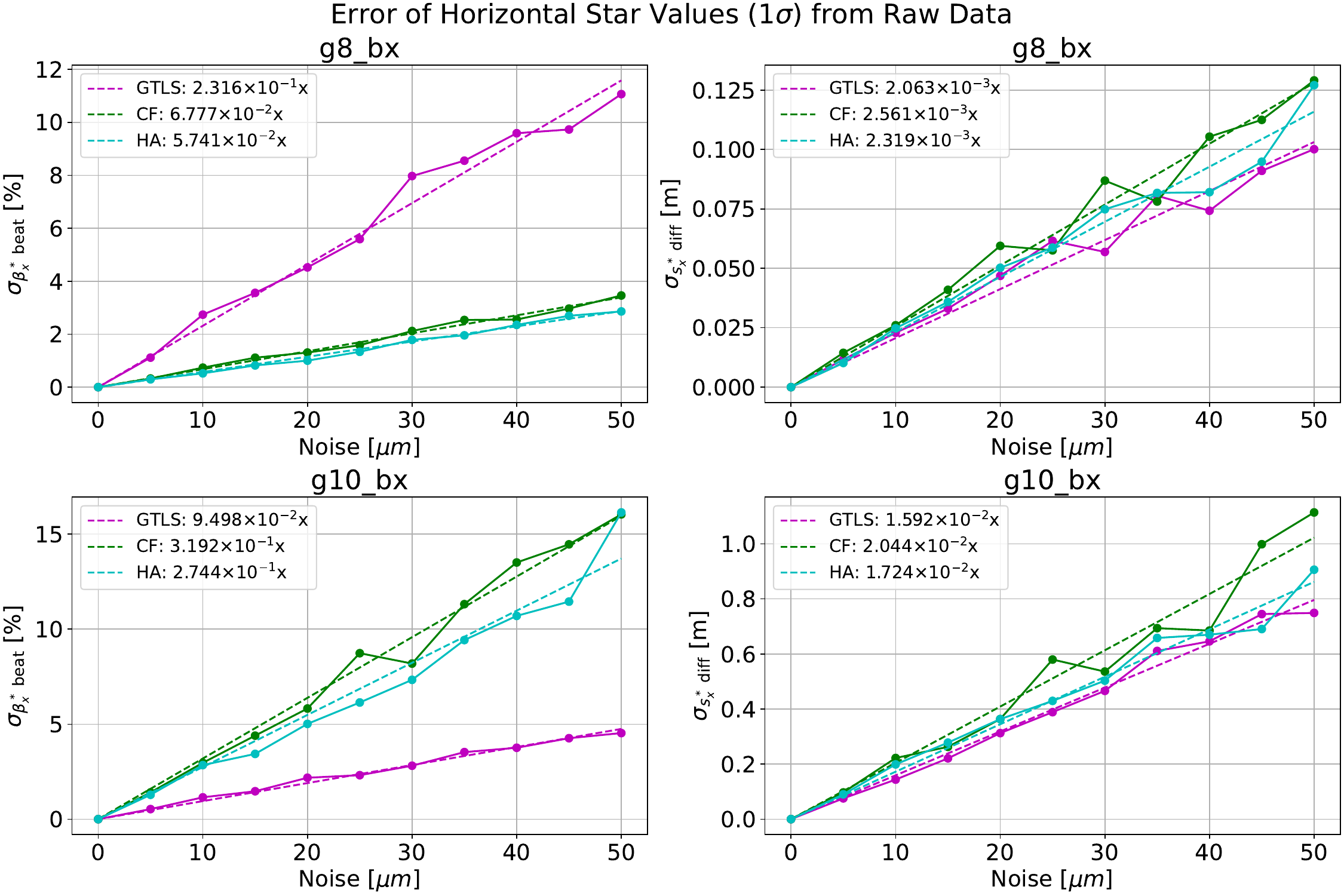}
    \caption{\label{fig:stdv_star} 
    Error trends of GTLS (magenta), CF (green), and HA (cyan) vs BPM error level from 0 to 50$\mathrm{\mu m}$ for $\beta^*$ beat (left) and $s^*$ difference (right) for IP8 (top) and IP10 (bottom) were calculated using Equation~(\ref{beta at IR}). The points represent the standard deviation values used for the linear regression calculation of the error trends, where $x$ in the legend is the noise level.}
\end{figure*}

The standard deviations from the Monte Carlo method for the calculation of $\beta^*$ and $s^*$ are given by Figure~\ref{fig:stdv_star}. A similar linear regression procedure was done here as well. The figure shows that lower errors from CF and HA at IP8 and lower errors from LR methods at IP10. However, the error is about the same in $s^*$ at both IPs.

These results can be explained by looking at an analytical calculation of $\beta^* \text{ and } s^*$. From Equation~(\ref{beta at IR}), the following forms for $\beta^*$ and $s^*$ can be calculated:

\begin{subequations}
    \begin{align}
        \beta^* = \frac{(\Delta s)^2 (\beta_1 + \beta_2 - 2Q)}{D} = \frac{N}{D}
    \end{align}
    \begin{align}
        s^* = \frac{s_1 + s_2}{2} - \frac{\Delta\beta\beta^*}{2\Delta s}
    \end{align}
\end{subequations}

\noindent where $\Delta s = s_1 - s_2, \Delta \beta = \beta_1 - \beta_2, N = (\Delta s)^2 (\beta_1 + \beta_2 - 2Q), D = (\Delta \beta)^2 + 4(\Delta s)^2, \text{ and } Q = \pm\sqrt{\beta_1\beta_2 - (\Delta s)^2}$ (the plus branch is selected for the physical solution). From these, the following errors can be calculated:

\begin{subequations}
    \begin{align}
        \frac{\partial \beta^*}{\partial \beta_i} = \frac{1}{D^2}(\frac{\partial N}{\partial \beta_i}D \mp 2N\Delta \beta)
    \end{align}
    \begin{align}
        \frac{\partial s^*}{\partial \beta_i} = \mp \frac{1}{2\Delta s}(\beta^* \pm \Delta \beta \frac{\partial \beta^*}{\partial \beta_i})
    \end{align}
\end{subequations}

\noindent where $i = 1, 2$ for the upstream and downstream BPMs. For symmetric beta at IP8, $\Delta \beta \rightarrow 0,  \beta_1 \approx \beta_2 \approx \beta_0, \text{ and } s_1 + s_2 = 0$. From Figure~\ref{fig:stdv_beta},  $\sigma_{\beta_1} \approx \sigma_{\beta_2} \approx \sigma_{\beta_0}$ for any measurement method. $\sigma_{\beta^*} \text{ and } \sigma_{s^*}$ can then be approximated as:

\begin{subequations}
    \begin{align}
        \sigma_{\beta^*} \approx (1 + \frac{\beta_0}{Q_0})\sigma_{\beta_0}\sqrt{\frac{1 + \rho}{8}}
    \end{align}
    \begin{align}
        \sigma_{s^*} \approx \frac{\beta^*}{2\Delta s}\sigma_{\beta_0}\sqrt{2(1 - \rho)}
    \end{align}
\end{subequations}

\noindent where $Q_0 = \sqrt{\beta_0^2 - (\Delta s)^2}$ and $\rho$ is the correlation coefficient between $\beta_1 \text{ and } \beta_2$. The analytical error propagation demonstrate that $\sigma_{\beta}$ from both BPMs scale linearly with both $\sigma_{\beta^*} \text{ and } \sigma_{s^*}$. This is all in accordance to Figure~\ref{fig:stdv_beta} and Figure~\ref{fig:stdv_star}. The discrepancy between methods in Figure~\ref{fig:stdv_star} can be explained by $\rho$, since $\rho$ is different for each optics measurement method and for different IPs (at IP8, $\rho \approx 0$ for CF/HA, $\rho \approx .7$ for OLS, and $\rho \approx .9$ for GTLS). $\beta_1$ and $\beta_2$ measured from the LR methods are both correlated though Equation~(\ref{angle coordinate}), using the same angle coordinate in their calculation. Furthermore, the SVD step in GTLS add to the positive correlation between $\beta_1$ and $\beta_2$. In contrast, $\beta_1$ and $\beta_2$ measured from CF and HA yield no correlation due to explicit normalization of the action with minimal parameters to be estimated. Due to this, the factor of $\sqrt{1 \pm \rho}$ drives estimates of the error up for $\sigma_{beta^*}$ and down for $\sigma_{s^*}$. Since $s^*$ primarily depends on $\beta_1 - \beta_2$, a strong positive correlation suppresses $\mathrm{Var}(\Delta\beta)$ within the LR methods, resulting in similar $\sigma_{s^*}$ across methods.

For non-symmetric $\beta$ at IP (or when $\Delta s^* \neq 0$), analytical calculations of star errors showed minimal changes in $\sigma_{\beta^*} \text{ and } \sigma_{s^*}$ ($\Delta \sigma_{\beta^*} \approx 0.001 \text{ and } \Delta \sigma_{s^*} \approx 0.01$ for $\Delta s^* = 0m \rightarrow \Delta s^* = .5m$). This was checked with Monte Carlo simulations in Xsuite where the lattice was changed to match the currents $\Delta I$ used in the ORM. This justifies averaging the $\beta^*$ beat and $s^*$ difference in Figure~\ref{fig:LOMM_IP8_err}.

\begin{figure*}[htbp]
    \centering
    \includegraphics[width=1.5\columnwidth]{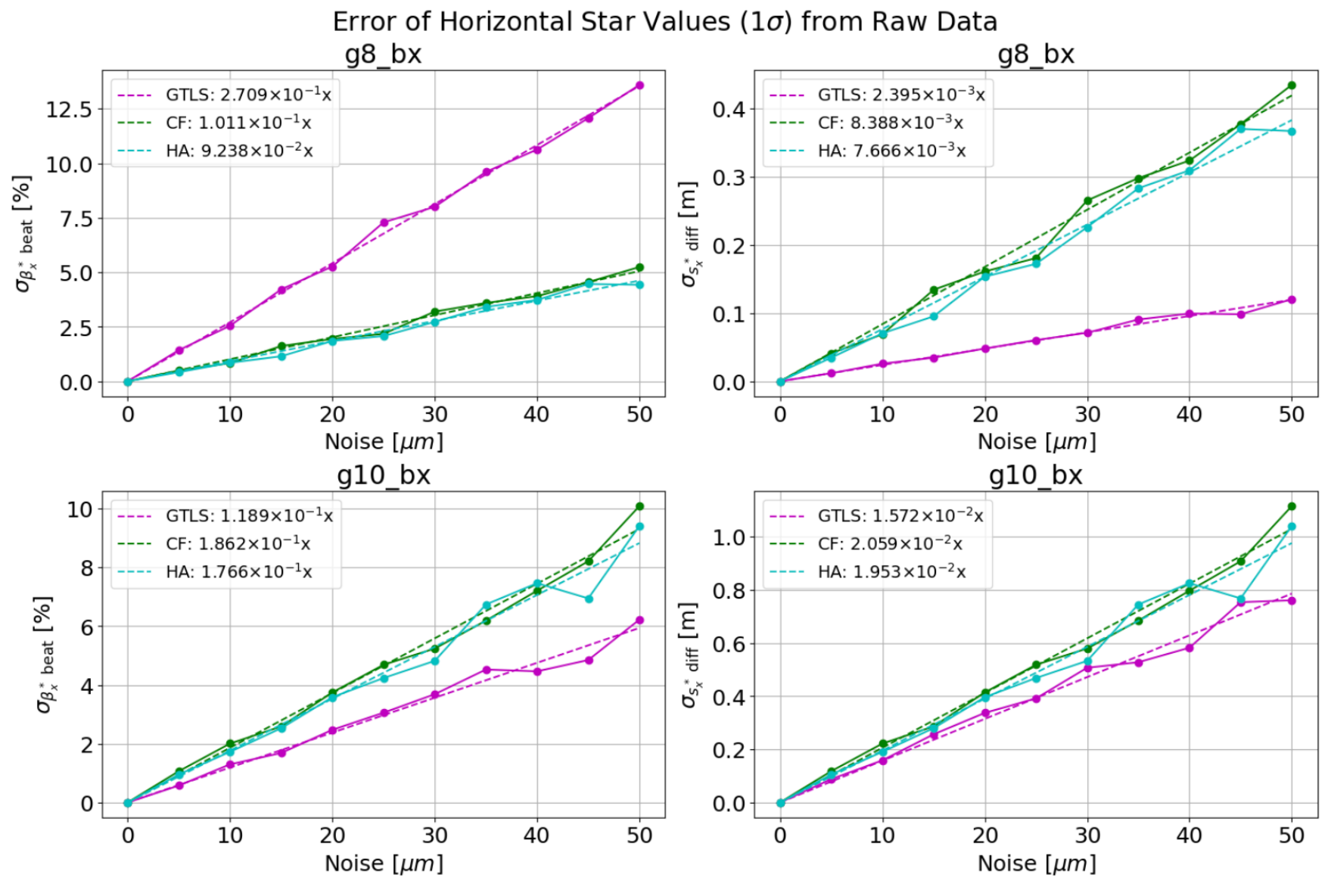}
    \caption{\label{fig:stdv_star_alpha} 
    Error trends of GTLS (magenta), CF (green), and HA (cyan) vs BPM error level from 0 to 50$\mathrm{\mu m}$ for $\beta^*$ beat (left) and $s^*$ difference (right) for IP8 (top) and IP10 (bottom) were calculated using Equation~(\ref{alpha at IR}). The points represent the standard deviation values used for the linear regression calculation of the error trends, where $x$ in the legend is the noise level.}
\end{figure*}

A complementary error propagation applies when $\beta^*$ and $s^*$ are calculated using Equation~(\ref{alpha at IR}) rather than Equation~(\ref{beta at IR}) as shown in Figure~\ref{fig:stdv_star_alpha}. When a middle-layer $\alpha$ calculation is introduced, since $\beta^*$ and $s^*$ calculations both depend on $\alpha$, all three methods experience an expected increase in error when compared to Figure~\ref{fig:stdv_star}. However, CF and HA experience a greater error amplification than GTLS. This is most visible in measurements at high beta regions.


\subsubsection{\label{sec: Gain Analysis} Gain Analysis}

\begin{figure*}[htbp]
    \centering
    \includegraphics[width=1.5\columnwidth]{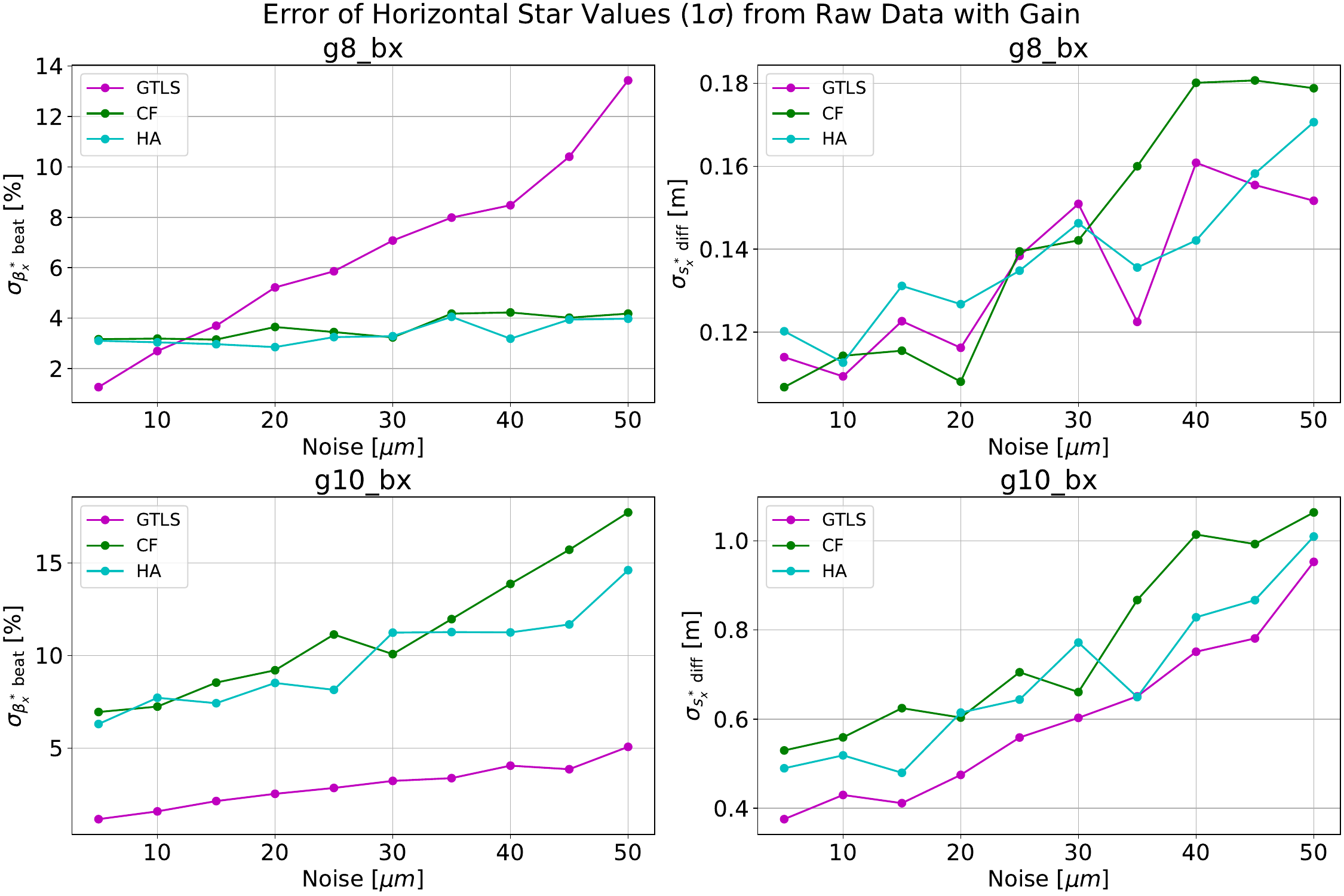}
    \caption{\label{fig:stdv_star_gain} 
    Error trends of GTLS (magenta), CF (green), and HA (cyan) vs BPM error level from 5 to 50$\mathrm{\mu m}$ and a constant gain of 2\% if noise was added for $\beta^*$ beat (left) and $s^*$ difference (right) for IP8 (top) and IP10 (bottom). The points represent the standard deviation values used for the linear regression calculation of the error trends, where $x$ in the legend is the noise level.}
\end{figure*}

An additional Monte Carlo simulation was ran for a noise variation involving BPM gain errors. Gain error was added in accordance with \cite{Safranek}:

\begin{equation}\label{gain}
    u_{meas} = (1 + \delta g)u_{true} + \sigma_{u}
\end{equation}

BPM calibration error at RHIC are usually measured to be within $2\%$ \cite{ICARHIC}. Since gain makes up a fraction of the calibration error, we estimate the gain to also be $2\%$ for the maximum potential error. Applying this to Equation~(\ref{gain}), Figure~\ref{fig:stdv_star_gain} is obtained for high and low beta regions. Since a constant gain of $2\%$ was applied, the point corresponding to zero random noise $(\sigma_u = 0)$ will not be error-free.

When comparing Figure~\ref{fig:stdv_star_gain} to Figure~\ref{fig:stdv_star}, this addition affects the error of the $\beta^*$ beat for CF and HA more so than the LR methods. This is as expected; the calculation of the action in CF and HA methods is affected by the gain. However it increases the error for measuring $s^*$ for all methods by a slight amount. Large gain errors possibly lead to beta functions that are not symmetric at IRs, making $s^*$ values large, regardless of method. Therefore, gain and other calibration errors should be minimized if possible.




\bibliographystyle{elsarticle-num} 
\bibliography{example}

@inproceedings{lum,
    title={Concept of luminosity},
    booktitle    = {CERN Accelerator School: Intermediate Course on Accelerator Physics},
    author={Herr, Werner and Muratori, Bruno},
    year={2006},
    publisher={CERN},
    pages={140–146},
    doi = {10.1007/978-3-642-23053-0\_9},
}

@manual{madxmanual,
  title        = {{The MAD-X Program (Methodical Accelerator Design), Version 5.08.01: User’s Reference Manual}},
  author       = {Deniau, Laurent and Grote, Hans and Roy, Ghislain and Schmidt, Frank},
  organization = {European Organization for Nuclear Research (CERN)},
  year         = {2024},
  note         = {Editor: Laurent Deniau}
}

@inproceedings{madxmatching,
  author       = {R. de Maria and F. Schmidt and P. K. Skowronski},
  title        = {{Advances in Matching with MAD-X}},
  booktitle    = {Proceedings of the International Computational Accelerator Physics Conference (ICAP)},
  year         = {2006},
  pages        = {213--215},
  month        = sep,
  note         = {CAP06-TUPLT041},
  address      = {Chamonix, France}
}

@article{snakes,
  author       = {C. Liu and J. Kewisch and H. Huang and M. Minty},
  title        = {{Minimization of spin tune spread for preservation of spin polarization at RHIC}},
  journal      = {Phys. Rev. Accel. Beams},
  volume       = {22},
  number       = {6},
  pages        = {061002},
  year         = {2019},
  month        = {Jun},
  publisher    = {American Physical Society},
  DOI={10.1103/physrevaccelbeams.22.061002}
}

@article{review, 
    title={{Review of Linear Optics Measurement and correction for charged particle accelerators}}, 
    volume={20}, 
    DOI={10.1103/physrevaccelbeams.20.054801}, 
    number={5}, 
    journal={Physical Review Accelerators and Beams}, 
    author={Tom{\'a}s, Rogelio and Aiba, Masamitsu and Franchi, Andrea and Iriso, Ubaldo}, 
    year={2017}, 
    month=May
}

@phdthesis{Castro,
  author       = {Castro, P.},
  title        = {{Luminosity and Beta Function Measurement at the Electron-Positron Collider Ring LEP}},
  school       = {CERN},
  year         = {1996},
  type         = {Ph{D} thesis}
}

@article{MIA,
  author       = {Wang, Chun and Sajaev, Vadim and Yao, Chih},
  title        = {{Phase advance and $\beta$-function measurements using model-independent analysis}},
  journal      = {Phys. Rev. Accel. Beams},
  volume       = {6},
  pages        = {104001},
  year         = {2003},
  doi          = {10.1103/PhysRevSTAB.6.104001},
  publisher    = {American Physical Society}
}

@article{NBPM,
  author       = {Langner, A. and Benedetti, G. and Carl{\'a}, M. and Iriso, U. and Mart{\'i}, Z. and Coello de Portugal, J. and Tom{\'a}s, R.},
  title        = {{Utilizing the ${N}$ beam position monitor method for turn-by-turn optics measurements}},
  journal      = {Phys. Rev. Accel. Beams},
  volume       = {19},
  number       = {9},
  pages        = {092803},
  year         = {2016},
  doi          = {10.1103/PhysRevAccelBeams.19.092803},
  publisher    = {American Physical Society},
  note         = {(Received 20 March 2016; published 21 September 2016)}
}

@article{NBPM2,
  author       = {Wegscheider, A. and Langner, A. and Tom{\'a}s, R. and Franchi, A.},
  title        = {{Analytical N-beam position monitor method}},
  journal      = {Phys. Rev. Accel. Beams},
  volume       = {20},
  number       = {11},
  pages        = {111002},
  year         = {2017},
  publisher    = {American Physical Society},
  doi          = {10.1103/PhysRevAccelBeams.20.111002},
}

@techreport{Decoherence,
  author       = {Meller, R. E.},
  title        = {{Decoherence of Kicked Beams}},
  institution  = {Fermilab},
  year         = {1987}
}

@article{ICARHIC,
  author       = {Shen, X. and Lee, S. Y. and Bai, M. and White, S. and Robert-Demolaize, G. and Luo, Y. and Marusic, A. and Tom{\'a}s, R.},
  title        = {{Application of independent component analysis to AC dipole based optics measurement and correction at the Relativistic Heavy Ion Collider}},
  journal      = {Phys. Rev. Accel. Beams},
  volume       = {16},
  number       = {11},
  pages        = {111001},
  year         = {2013},
  month        = nov,
  doi          = {10.1103/PhysRevSTAB.16.111001}
}

@article{sPHENIX,
  author       = {Campbell, Sarah},
  title        = {{sPHENIX: The Next Generation Heavy Ion Detector at RHIC}},
  journal      = {Journal of Physics: Conference Series},
  year         = {2017},
  DOI={10.1088/1742-6596/832/1/012012}
}

@article{coupling,
  author       = {W. Fischer},
  title        = {{Robust linear coupling correction with N-turn maps}},
  journal      = {Phys. Rev. ST Accel. Beams},
  volume       = {6},
  pages        = {062801},
  year         = {2003},
  doi          = {10.1103/PhysRevSTAB.6.062801}
}

@book{Wiedemann,
  author       = {Wiedemann, Helmut},
  title        = {{Particle Accelerator Physics}},
  edition      = {Fourth},
  publisher    = {Springer Nature},
  year         = {2015},
  pages        = {213--231}
}

@techreport{Liu,
  author       = {Liu, C.},
  title        = {{Optics Measurement and Correction During Beam Acceleration in the Relativistic Heavy Ion Collider}},
  institution  = {Brookhaven National Laboratory (BNL)},
  address      = {Upton, NY, United States},
  month        = sep,
  year         = {2014},
  DOI={10.2172/1159698}
}

@techreport{RHICConfigManual,
  author       = {Collider-Accelerator-Department},
  title        = {{RHIC Configuration Manual}},
  institution  = {Brookhaven National Laboratory},
  year         = {2006},
}

@article{Syphers,
  author       = {Syphers, M. J. and Miyamoto, R.},
  title        = {{Direct Measurements of Beta-Star in the Tevatron}},
  journal      = {IEEE Transactions on Nuclear Science},
  year         = {2007},
  DOI={10.1109/pac.2007.4440470}
}

@incollection{Minty,
  author       = {Minty, Michiko G. and Zimmermann, Frank},
  title        = {{Transverse Optics Measurement and Correction: Multiknobs, Optics Tuning, and Monitoring}},
  booktitle    = {Measurement and Control of Charged Particle Beams},
  publisher    = {Springer Nature},
  year         = {2003},
  pages        = {43--46},
  DOI={10.1007/978-3-662-08581-3\_2}
}

@article{TLS, 
 title={{An analysis of the total least squares problem}}, 
 volume={17}, 
 DOI={10.1137/0717073}, 
 number={6}, 
 journal={SIAM Journal on Numerical Analysis}, 
 author={Golub, Gene H. and van Loan, Charles F.}, 
 year={1980}, 
 month=Dec, 
 pages={883–893}}

@book{LR,
  author    = {Greene, William H.},
  title     = {{Econometric Analysis}},
  edition   = {5},
  year      = {2003},
  publisher = {Prentice Hall},
  address   = {Upper Saddle River, NJ},
  isbn      = {978-0-13-066189-0}
}

@article{GTLS,
title = {{Overview of total least-squares methods}},
journal = {Signal Processing},
volume = {87},
number = {10},
pages = {2283-2302},
year = {2007},
note = {Special Section: Total Least Squares and Errors-in-Variables Modeling},
issn = {0165-1684},
doi = {https://doi.org/10.1016/j.sigpro.2007.04.004},
author = {Ivan Markovsky and Sabine {Van Huffel}}}

@article{Edwards_Teng, 
 title={{Parametrization of linear coupled motion in periodic systems}}, 
 volume={20}, 
 DOI={10.1109/tns.1973.4327279}, 
 number={3}, 
 journal={IEEE Transactions on Nuclear Science}, 
 author={Edwards, D. A. and Teng, L. C.}, 
 year={1973}, 
 pages={885–888}}

@ARTICLE{Xsuite,
  author  = {G. Iadarola and R. De Maria and S. {\L}opaciuk and A. Abramov and
             X. Buffat and D. Demetriadou and L. Deniau and P. Hermes and
             P. Kicsiny and P. Kruyt and A. Latina and L. Mether and
             K. Paraschou and G. Sterbini and F. F. Van Der Veken and
             P. Belanger and P. Niedermayer and D. Di Croce and T. Pieloni and
             L. Van Riesen-Haupt and M. Seidel},
  title   = {{Xsuite: An Integrated Beam Physics Simulation Framework}},
  journal = {JACoW HB2023},
  year    = {2024},
  pages   = {TUA2I1},
}

@techreport{Luo_RHIC_DynamicAperture_2009,
  author      = {Y. Luo},
  title       = {{Dynamic Aperture Evaluation of the Proposed Lattices for the 2009 Polarized Proton Runs}},
  institution = {U.S. Department of Energy / Brookhaven National Laboratory},
  year        = {2009},
  doi         = {doi.org/10.2172/946785}
}

@article{Safranek,
  author       = {Safranek, J.},
  title        = {{Experimental Determination of Storage Ring Optics Using Orbit Response Measurements}},
  journal      = {Nuclear Instruments and Methods in Physics Research Section A},
  year         = {1997},
  volume       = {388},
  pages        = {27--36},
  DOI={10.1016/s0168-9002(97)00309-4}
}

@article{PhysRevAccelBeams.23.042801,
  title = {{Optics-measurement-based beam position monitor calibrations in the LHC insertion regions}},
  author = {Valdivieso, A. Garc\'{\i}a-Tabar\'es and Tom\'as, R.},
  journal = {Phys. Rev. Accel. Beams},
  volume = {23},
  issue = {4},
  pages = {042801},
  numpages = {15},
  year = {2020},
  month = {Apr},
  publisher = {American Physical Society},
  doi = {10.1103/PhysRevAccelBeams.23.042801},
}

@article{Carlier2017BetaStar,
  author    = {Carlier, F. and Tom{\'a}s, R.},
  title     = {{Accuracy and feasibility of the $\beta^*$ measurement for LHC and High Luminosity LHC using k modulation}},
  journal   = {Physical Review Accelerators and Beams},
  volume    = {20},
  number    = {1},
  pages     = {011005},
  year      = {2017},
  doi       = {10.1103/PhysRevAccelBeams.20.011005},
  publisher = {American Physical Society}
}






\end{document}